\documentclass[aps,prb,twocolumn,groupedaddress,showpacs,amssymb,amsmath,letterpaper]{revtex4}
\usepackage{graphicx}
\usepackage[final,dvips]{epsfig}
\usepackage{color}
\usepackage{dcolumn}
\usepackage{bm}
\usepackage{array}
\usepackage{amssymb}

\begin{document}

\title{Spatially inhomogeneous phase in the two-dimensional repulsive Hubbard model}

\author{Chia-Chen Chang}

\author{Shiwei Zhang}

\affiliation{Department of Physics, College of William and Mary, Williamsburg, VA 23187}

\begin{abstract}   
  Using recent advances in auxiliary-field quantum Monte Carlo techniques and
  the phaseless approximation to control the sign/phase problem, we
  determine the equation of state in the ground state of the two-dimensional
  repulsive single-band Hubbard model at intermediate interactions.
  Shell effects are eliminated and finite-size effects are greatly reduced by boundary condition
  integration.  Spin-spin correlation functions and structure factors are also
  calculated.  In lattice sizes up to $16\times 16$, the results show signal for phase-separation.  Upon
  doping, the system separates into one phase of density $n=1$ (hole-free)
  and the other at density $n_c$ ($\sim 0.9$).  The long-range
  antiferromagnetic order is coupled to this process, and is lost below
  $n_c$.
\end{abstract}

\pacs{71.10.Fd, 02.70.Ss}

\maketitle

\section{Introduction}

The Hubbard model
\cite{Hubbard}
provides a minimal framework for describing electron
interactions in a crystal lattice,
and has played a central role in condensed matter and
quantum many-body physics.
Especially since the discovery of
high-$T_c$ superconductors,
the two-dimensional (2-D) Hubbard model,
believed to contain the essential
physics of the CuO plane
\cite{Anderson1987},
has been intensely studied.
The combination of theoretical and numerical techniques has made
important progress \cite{Dagotto1984,Scalapino2004},
 but some basic questions 
have remained.

One of the questions is 
whether there is phase separation (PS) in the ground state
of the Hubbard model.  
The question is important in its own right, as
a key element in our understanding of the 
phase diagram of 
this fundamental model. 
Recent experimental indication of spatial 
inhomogeneities in cuprates \cite{Charge}
has further increased its potential 
relevance 
and interest.
In the past two decades a large body of numerical work has been devoted 
to resolving this
issue
\cite{Emery1990,StephenHellberg1997,Moreo1991,
Lin1991,Becca2000,Cosentini1998,Zitzler2002,Macridin2006,Aichhorn2007},
but the results have been conflicting.
The differing answers underscore the challenges:
the requirement of
high accuracy, as well as the difficulty in extrapolating to the 
thermodynamic limit
because of extreme sensitivity of the signal to both finite-size and
shell effects.

In this paper, we apply recent advances
in auxiliary-field quantum Monte Carlo (QMC) techniques \cite{Zhang1997,Zhang2003}
to study the ground state of the repulsive 2-D Hubbard model.
Our goal was to shed light on the question of PS. 
A second motivation comes from ultra-cold atoms, where
rapid experimental progress
promises a new avenue ---
optical-lattice emulators \cite{Emulator} ---
for direct ``simulations'' to
investigate properties of Hubbard-like models.
Detailed, accurate numerical data would allow
quantitative benchmark and comparisons in future
optical-lattice experiments.
In our approach, the ability to control the sign/phase problem with a good 
approximation, combined with
a boundary condition integration technique, drastically reduces the
finite size and shell effects.
This allows us to reach 
much higher accuracy than previously possible in the model.
The measured equation of state 
and spin-spin correlations, in lattice sizes up to $16\times 16$, show clear
signals for PS 
at intermediate
interaction strengths.
The nature of this spatially inhomogeneous state is examined.


The Hamiltonian for the
one-band Hubbard model is:
\begin{eqnarray}
  H &=& -t \sum_{\mathbf j, \bm\delta,\sigma} 
        \left(  c_{\mathbf j,\sigma}^\dagger c_{\mathbf j+\bm\delta,\sigma} +\mbox{h.c.} \right)
      + U\sum_{\mathbf j} n_{\mathbf j\uparrow} n_{\mathbf j\downarrow},
\label{Hamiltonian}
\end{eqnarray}
where $c_{\mathbf{j},\sigma}^\dagger$ ($c_{\mathbf{j},\sigma}$) creates (annihilates) an electron with spin $\sigma$ 
($\sigma=\uparrow, \downarrow$) at lattice site $\mathbf{j}$,  
and $\bm\delta$ connects two nearest-neighbor
sites.
The square lattice has size $N=L\times L$, with
$N_\sigma$ spin-$\sigma$ electrons.
The model has only two parameters, the strength of the
interaction $U/t$ (we will set $t=1$) and the
electron density $n\equiv (N_\uparrow+N_\downarrow)/N$.

PS occurs when the stability 
condition $\partial^2 e(n)/\partial n^2 > 0$ is violated, 
where $e(n)$ is the ground-state energy (per site)
at density $n$. The critical 
value of $n$ 
can be identified by Maxwell construction.
Emery {\it et al.}~\cite{Emery1990} showed that in the Hubbard (or $t$-$J$) model one
could study
\begin{equation}
 e_h(h)\equiv\frac{e(1-h)-e(1)}{h}, \label{HoleEnergy}
\end{equation}
where $h$ is the hole density: $h \equiv 1-n$. 
If PS exists, there is a minimum in $e_h(h)$ at $h_c$ 
(or in the thermodynamic limit,
a constant $e_h(h)$ for $h<h_c$) \cite{Emery1990,Cosentini1998}. 
 
\section{Method}
\subsection{Twist-Averaged Boundary Condition (TABC)}

\begin{figure}
\includegraphics[width=0.68\columnwidth,angle=-90]{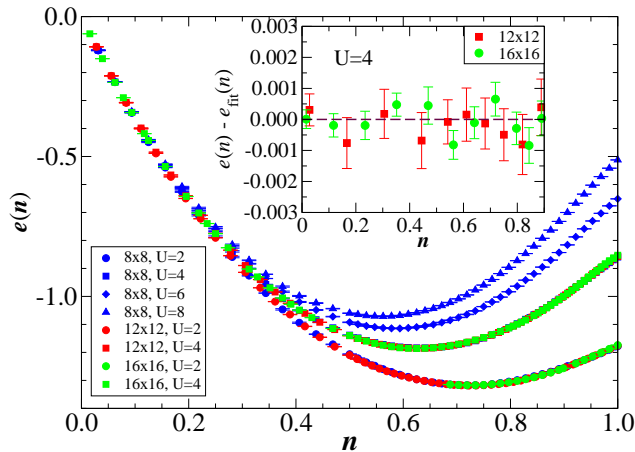} 
\caption{
(Color online) Ground state energy per site of the 2-D Hubbard model vs.~density, 
for several interaction strengths and
lattice sizes. 
Error bars
are combined QMC and $\mathbf{\Theta}$-integration statistical 
errors. 
As a result of TABC, 
curves are smooth and 
different lattice sizes are indistinguishable. 
The inset shows convergence to the thermodynamic limit with 
a magnified view.
(To reduce clutter, only every fifth density is shown for each size.)
It also illustrates the accuracy of the fit $e_{\rm fit}(n)$ 
across the density range
for the phase below $n_c$. 
}
\label{Egr}
\end{figure}

The signal for PS from Eq.~(\ref{HoleEnergy})
requires the slope of the equation of state, i.e., accurate numerical  
determination of small energy differences in the region where
$h$ is small.
For a finite lattice, the shape of the Fermi surface varies considerably 
with $n$, which causes large variations in the energy.
For example, with the usual periodic boundary 
condition (PBC),
the smallest $h$ accessible by a closed-shell 
system
is $\sim 0.15$ 
in a $16\times 16$ lattice \cite{note_tilt};
even at $40\times 40$ the finite-size effect is still sizable, especially
in the region relavant for PS
(see inset in Fig.~\ref{Ehole}). 
To reduce shell and finite-size effects, 
we use twist-averaged boundary condition
(TABC)\cite{Poilblanc1991,Gross,Lin2001}, under which
the wave function $\Psi(\mathbf{r}_1,\mathbf{r}_2,\ldots)$ gains a phase when
electrons hop around lattice boundaries:
\begin{equation}
  \Psi(\ldots,\mathbf{r}_j+\mathbf{L},\ldots) = e^{i\widehat{\mathbf{L}}\cdot\mathbf{\Theta}}\Psi(\ldots,\mathbf{r}_j,\ldots),
\end{equation}
where $\widehat{\mathbf{L}}$ is the unit vector along $\mathbf{L}$,
and  
the 
twist angle $\mathbf{\Theta}=(\theta_x,\theta_y)$ is a
parameter. 
With a generic $\mathbf{\Theta}$, there will be no degeneracy in the 
one-electron energy levels.
We average the results over many
random twist angles \cite{Lin2001} in each system
for convergence.
As shown in Figs.~\ref{Egr} and~\ref{Ehole}, TABC 
essentially eliminates any
shell effect.
The disadvantage is that it turns the QMC sign problem \cite{Zhang1997}
into a phase problem \cite{Zhang2003}.

\begin{figure}
\includegraphics[width=0.68\columnwidth,angle=-90]{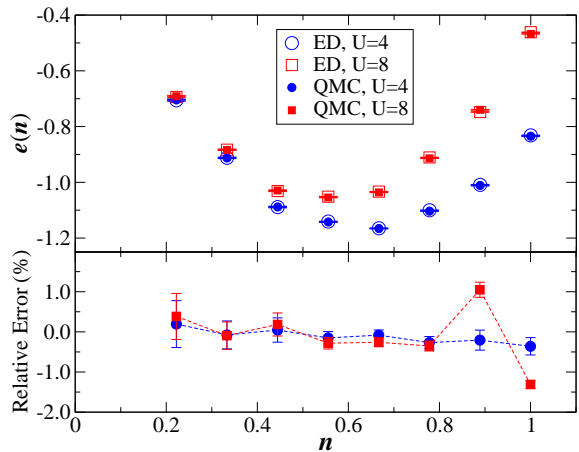} 
\caption{
(Color online) Upper panel: Ground state energy per site $e(n)$, versus
density, of 
the $3\times 3$ 
Hubbard lattice at $U=4$ (blue) and $8$ (red) calculated by ED 
(empty symbols) and our QMC method (filled symbols). At each density,
the result is the average from $1000$ random 
$\mathbf{\Theta}$ values and the 
statistical error is estimated from their distribution.
Bottom panel: Relative error (see text) 
of QMC ground state energy 
compared to the exact result  
(percentage).
}
\label{Exact.vs.QMC}
\end{figure}

\subsection{Constrained Path Monte Carlo under TABC}

To treat this problem, we extend the 
constrained path Monte Carlo (CPMC) method
\cite{Zhang1997} to 
a Hamiltonian
under TABC.
For each given system (specified by $N$, $n$, $U$, and $\mathbf{\Theta}$),
the method obtains
a Monte Carlo (MC) representation of 
the many-body ground state $|\Psi_G\rangle$,
by importance-sampled branching random walks (RWs) 
\cite{Zhang1997,Zhang2003}
in the space of 
Slater determinant wave functions.
The usual sign problem under PBC is caused by the
symmetry \cite{Zhang1991,Zhang1997}
between a Slater determinant $|\phi\rangle$ 
and a degenerate partner $-|\phi\rangle$ (exchanging two 
orbitals). To specify $|\Psi_G\rangle$, we need either, 
but not both.
It can be shown \cite{Zhang1997,Zhang1999} that 
constraining the RWs to 
$\langle \Psi_G|\phi\rangle>0 $ is an exact boundary condition
that eliminates the sign problem.
In the constrained path approximation, a trial wave function 
$|\Psi_T\rangle$ is used in place of $|\Psi_G\rangle$.

Under TABC, the Slater determinants become complex, and
we need
to break the phase symmetry in $|\phi\rangle$.
The Hubbard-Strotonivich transformation
used in our calculations is
the spin-decomposition of Hirsch \cite{Hirsch1985},
which results in {\em real\/} Ising-like auxiliary fields.
The phase problem comes only from 
one-body hopping terms.
We use a simple version of the phaseless approximation \cite{Zhang2003}   
to constrain $|\phi\rangle$ to a unique phase. 
At each step of propagation, the paths of the RWs are required to 
satisfy:
\begin{equation}
 \Re\left\{\frac{\langle\Psi_T|\phi'\rangle}{\langle\Psi_T|\phi\rangle}\right\} > 0,
\label{eq:constraint}
\end{equation}
where $|\phi\rangle$ and 
$|\phi'\rangle$ are the current and proposed positions.
The left-hand side is used in the importance sampling
\cite{Zhang1997,Zhang2003,Purwanto2004}.
We use the free-electron wave function as 
$|\Psi_T\rangle$.
Since this is an eigenfunction of the complex
kinetic energy terms of $H$,
all the phase effect is absorbed in the {\em deterministic\/} one-body part.
The condition on the RWs is equivalent to the 
original constrained path approximation\cite{Zhang1997},
to which Eq.~(\ref{eq:constraint}) reduces if
$\mathbf{\Theta}=0$.
The phase constraint in
Eq.~(\ref{eq:constraint}) is the only approximation in our method.

Since the approximation involves only 
the overall 
sign/phase of the many-body wave function,
it is reasonable to expect that the results will be relatively 
insensitive to $|\Psi_T\rangle$.
Extensive 
benchmarks have shown this to be the case.
The general approach has,
in a variety of systems \cite{Zhang1997,AlSaidi2006,Suewattana2007,Kwee2008}, 
given results among the most accurate that can 
be achieved presently from QMC.

As a quantative measure in the current case, we 
compare
$e(n)$ in $3\times 3$ Hubbard lattices 
($U=4$ and $8$) between our method (QMC)
and
exact diagonalization (ED). 
At each density (both $N_\uparrow=N_\downarrow$ and the polarized 
case $N_\uparrow-N_\downarrow=1$, with $N_\downarrow=1,2,3,4$), 
we calculate the  ground-state energies for 
 1000 random $\mathbf{\Theta}$ values (identical 
in QMC and ED), average the results, and
estimate a statistical 
error bar.
In the QMC results, the error bar is the combined statistical errors
from the random 
$\mathbf{\Theta}$ distribution and the QMC sampling, although 
the latter is much smaller compared to the former in this system.
The results are shown in Fig.~\ref{Exact.vs.QMC}.
The agreement between QMC and exact results is excellent.
The relative error $[e_{\rm QMC}(n)-e_{\rm ED}(n)]/|e_{\rm ED}(n)|$, shown 
in the bottom panel, is essentially zero for $U=4$ and is less than $1.5\%$
for $U=8$, across the entire density range.  
 

\section{Results}
\subsection{Equation of state}

\begin{figure}
\includegraphics[width=0.7\columnwidth,angle=-90]{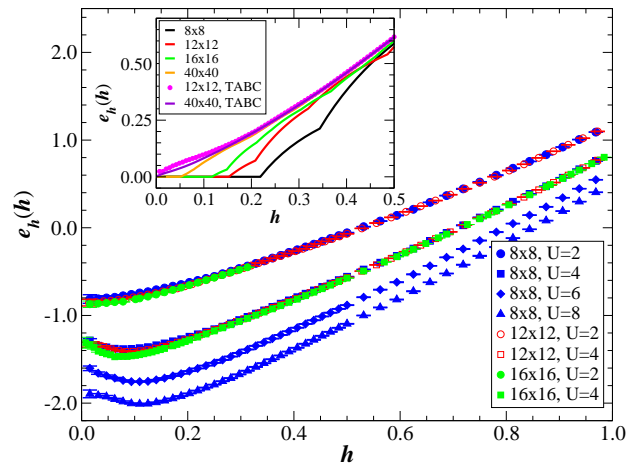} 
\caption{
(Color online) The hole energy $e_h(h)$ vs.~hole density $h$ for interacting systems,
derived from Fig.~\ref{Egr}.
A clear minimum is seen for  $U\geq 4$, at finite hole density $h_c$.
The inset shows $e_h(h)$
for {\it non}-interacting Hubbard model calculated for lattices up to $40\times 40$ 
with PBC.
Note the kinks and the flat part of the curves near half-filling. The magenta curve is for a
$12\times 12$ lattice (and the dashed line,
$40\times 40$) using TABC, which effectively eliminates
the finite-size and shell effects.
}
\label{Ehole}
\end{figure}

Our main energy results  
are summarized in Fig.'s~\ref{Egr} and \ref{Ehole}. In Fig.~\ref{Egr}, the
equation of state is presented 
for several lattice sizes and interaction strengths. 
For densities $n\lesssim 0.9$,
convergence of the averaged energy is rapid 
with respect to the set of random twists, and 
typically $20$ $\mathbf{\Theta}$'s is sufficient. 
For densities closer to half-filling, the energy has stronger fluctuations 
with $\mathbf{\Theta}$. Further, the requirement on statistical accuracy is 
higher in this region, because the error bar on $e_h(h)$
is magnified by $1/h$ 
(see Eq.~(\ref{HoleEnergy})).
In this case, the number of boundary
conditions is increased (to $60$-$300$). 
In each region, the same set of random $\mathbf{\Theta}$ values
are used to help correlate the results at different densities.
The main graph shows results from a Trotter time step 
$\Delta\tau=0.05$; the fit below [Eq.~(\ref{eq:efit})] 
and results in the inset have been 
extrapolated to $\Delta \tau=0$. 
Convergence to the thermodynamic limit is seen with all three lattice sizes
in the main graph.
As the inset shows,  
$12\times 12$ and $16\times 16$ are indistinguishable
to within statistical errors ($\sim 10^{-3}$).

In Fig.~\ref{Ehole}, the hole energy $e_h(h)$ derived from $e(n)$ 
is plotted. 
The inset illustrates the large finite-size and shell effects under the 
usual PBC. 
Because of degeneracies at the Fermi surface,
the hole energy 
has kinks and is a constant below 
a finite hole concentration\cite{Lin1991}.
As the system size is increased, the $e_h(h)$ curves 
show convergence, but only slowly.
Indeed
a false signal for PS is seen in the non-interacting systems. 
These features are removed by TABC, with which a smooth monotonic curve 
is obtained. 
Excellent convergence toward the thermodynamic limit
is achieved with a $12\times 12$ lattice.

Interacting systems 
show similar behaviors: 
under PBC the same kinks 
appear in the $e(n)$ vs.~$n$ curves \cite{Furukawa1992,Zhang1995}
for the interaction strengths considered here. 
The combination of CPMC and TABC leads to a dramatic improvement.
The main panel of Fig.~\ref{Ehole} shows the hole energy
for interacting systems.
A clear minimum in $e_h(h)$ can be seen at a finite hole density in all 
cases when $U\ge 4$. 
At $U=4$, $h_c$ is $\sim0.07$-$0.1$. 
As $U$ is
increased, the position of the minimum is seen to 
shift to the right, i.e., to a larger $h_c$.
As $U$ decreases to $U=2$, 
$e_h(h)$ appears to decrease
monotonically down to $h\sim 0.014$, the lowest doping  
in these lattices, although 
it cannot be completely ruled out
a shallow ($< 0.03$ from $e_h(0)$) minimum
exists
within the statistical error bars.

The energy results indicate that, near half-filling, the system
phase-separates into a hole-free phase of density $n=1$ and a
phase at $n_c=1-h_c$. 
Within a single phase, our results are expected to be at or near the 
thermodynamic limit. If the system is in a mixed state with two or 
more phases present, however, there are likely finite-size and/or interface
effects. 
This appears to be the case from the data, where
we see a minimum in the hole energy curves (as opposed to 
a flat region), as well as size variations in $e_h(h)$ in the 
hole density range $0<h\lesssim h_c$.
Similarly, if the system is in a spatially inhomogeneous 
spin or charge density wave state with very long wavelength modulations,
for example a stripelike state with only one stripe in a 
lattice of linear dimension up to $L\sim 16$, 
finite-size effects would likely make it indistinguishable 
from a phase-separated state in our calculations.

As a simple way to characterize the equation of state in 
the thermodynamic limit at $n<n_c$, we fit 
the calculated $e(n)$ on $n\in(0,0.9)$
 (size $L\ge 12$ only) to a 4-th order polynomial.
For $U=4$ this gives
\begin{equation}
e_{\rm fit}(n) = -4.004\,n + 3.769\,n^2 -0.700\,n^3 + 0.091\,n^4.
\label{eq:efit}
\end{equation}
Statistical errors in the fitted coefficients are 
$10^{-3}$ to $10^{-2}$. The inset in Fig.~\ref{Egr} shows
the quality of the fit.

\subsection{Spin-Spin Correlation}

\begin{figure}
\centering
\includegraphics[width=0.69\columnwidth,angle=-90]{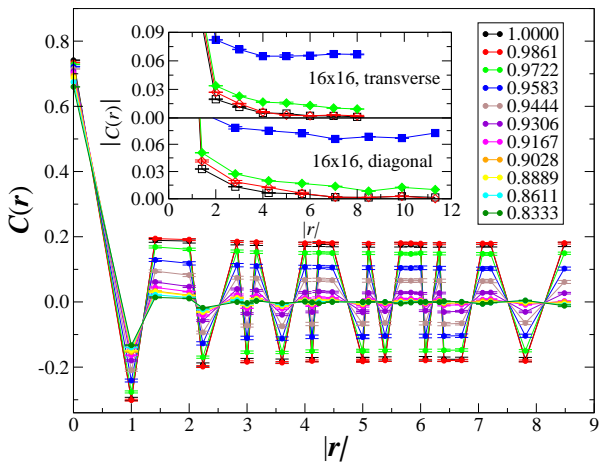}
\caption{
(Color online) Spin-spin correlation function $C({\bm r})$ for a $12\times 12$ Hubbard
lattice with $U=4$. 
Within the PS region,
the system exhibits long-range 
AF correlation. The strength of the long-range 
correlation decreases with doping, 
and vanishes at smaller densities.
The inset shows a $16\times 16$ lattice at $U=4$, at a few selected 
densities near $n_c$:
$n=0.9688$ (blue square), $0.9453$ (green diamond), $0.9297$ (red empty diamond), and $0.9063$ (black empty square).
To aid the eye, the absolute value $|C({\mathbf r})|$ is shown,
along two separate directions.
The behavior of the curves indicates 
the finite sizes of the AF phase in the periodic 
lattice.
}
\label{Spin}
\end{figure}

At $n=1$, the ground state is known to exhibit  
long-range antiferromagnetic (AF) order.\cite{Hirsch1985,Hirsch1989} 
Doping introduces frustration and tends to destroy the AF order.
To see how this occurs and the relation to PS,
we use the back-propagation 
technique\cite{Zhang1997,Purwanto2004} to calculate the spin-spin correlation function:
\begin{equation}
  C({\mathbf r})=\frac{1}{N}\sum_{{\mathbf j}}
  \langle
  (n_{{\mathbf j}+{\mathbf r},\uparrow}-n_{{\mathbf j}+{\mathbf r},\downarrow})(n_{{\mathbf j},\uparrow}-n_{{\mathbf j},\downarrow})
  \rangle,
\label{eq:defCr}
\end{equation}
where ${\mathbf r}$ is a vector on the lattice and $\langle..\rangle$ denotes
expectation with respect to the ground state.
The results for a $12\times 12$ lattice at $U=4$, after twist-averaging, 
are shown in Fig.~\ref{Spin}.
AF order is evident at $n=1$, as expected. 
Note that
the magnitude of the long-range part is $\sim 0.2$, and
double occupancy of $\uparrow$ and $\downarrow$-electrons is 
significant, as the strength of the interaction $U$ is moderate.
The long-range order decays rapidly with $n$ and,
in the homogeneous phase, only short range correlation remains.
(The minimum of $e_h(h)$ is around $n= 0.9167$ in $12\times 12$.) 

\begin{figure}
\centering
\includegraphics[width=0.68\columnwidth,angle=-90]{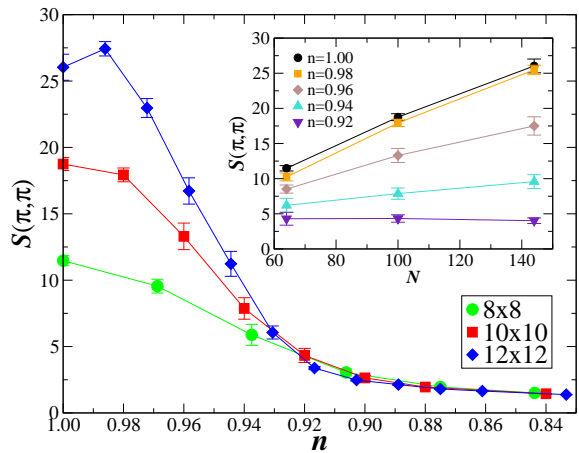} 
\caption{
(Color online) Spin structure factor at $\mathbf{q}=(\pi,\pi)$ for three system sizes
calculated at $U=4$. The lines are guides to the eye.
The inset shows $S(\pi,\pi)$  vs.~lattice size
at several densities (obtained by linear interpolation if 
the exact $n$ is not available in the particular lattice). 
}
\label{Sk}
\end{figure}

A more quantitative picture can be seen from 
the spin structure factor:
$S(\mathbf{q})=\sum_\mathbf{r} C(\mathbf{r})e^{i\mathbf{q}\cdot\mathbf{r}}$. 
When the system has
AF order, 
$S(\mathbf{q})$ will peak at $(\pi,\pi)$.
The calculated results are plotted in
Fig.~\ref{Sk}, as a 
function of $n$ for three different lattice sizes. 
There is a marked difference
between the small and larger doping regions.
Below a critical density
($n\lesssim n_c$),
$S(\pi,\pi)$ remains finite but is small and independent of
lattice size, indicating 
the presence of short-range spin correlation but no long-range magnetic
order. 
Beyond $n_c$,
$S(\pi,\pi)$ increases quickly
as $n$ approaches $1$. 
As the inset illustrates,
at each density
$S(\pi,\pi)$ grows proportionally with system size,
suggesting the presence of long-range AF order.

We now further examine the spatial dependence of the spin correlation.
From the Maxwell construction,
the size of 
the AF region in a phase-separated system ($n> n_c$) is 
$N_{\rm AF}=(1-h/h_c)\,N$.
In our calculations,
$C(\mathbf{r})$
is averaged over imaginary-time and 
MC configurations.
An AF cluster of 
linear dimension $l_{\rm AF}> L/2$ should,
due to ``winding'' around the periodic lattice, 
have a finite, constant tail $|C(\mathbf{r})|$ 
 beyond
$|\mathbf{r}|\sim L-l_{\rm AF}$, while
a smaller cluster should have a tail at zero
beyond $|\mathbf{r}| \sim l_{\rm AF}$.
Our $C(\mathbf{r})$ results are consistent with this.
In $12\times 12$, 
finite resolution gives only a handful of densities
on the interval $(n_c,n)$, so
$l_{\rm AF}$ is 
close to  either $L$ or $0$,
and we see long plateaus.
The inset in Fig.~\ref{Spin} shows $16\times 16$ lattices, focusing
on several densities near $n_c$.
At $n=0.9688$ and $0.9453$, $l_{\rm AF}>L/2$,
but the former (large $l_{\rm AF}$) has a long flat tail while the latter
shows a decline with $|\mathbf{r}|$ in the middle, indicating 
reduced contributions in the sum in Eq.~(\ref{eq:defCr}).
Similar effects are seen in the other pair ($l_{\rm AF}<L/2$),
with 
$n=0.9297$ showing an extended intermediate region in which 
$|C(\mathbf{r})|$ is finite but decreasing, before the vanishing tail.

\section{Discussion and Conclusion}

As we have discussed in Sec.~II.B, our calculations use a
non-perturbative,
many-body QMC method. 
We return again to the only approximation in the method, namely
the phase constraint, to help further 
gauge its impact.
Although the possibility of a systematic bias cannot be 
ruled out, every indication has been that
our results are very accurate ---
including the quality of the present data,
the consistency 
between the energy and spin correlation results, 
and the extensive benchmarks to date.
As mentioned, the constrained path approximation
has been tested (Refs.~\onlinecite{Zhang1997, Zhang1999, Zhang1995} and others) in various Hubbard  
systems under periodic or open boundary conditions.
Accurate energy results are obtained. 
In realistic electronic systems, an approximation which is based on the same
framework but which has to deal with a real two-body phase problem (as opposed
to the non-stochastic one-body hopping phase here) has been benchmarked 
in molecules (Refs.~\onlinecite{Zhang2003, Purwanto2004, AlSaidi2006, Suewattana2007} and others) against
density-matrix renormalization group and quantum chemistry methods.
Again the accuracy in the calculated ground state enerfy 
is consistent with that of Fig.~\ref{Exact.vs.QMC}.

In addition, several other factors in the present work provide more
self-consistency checks and 
show the robustness of the results.
At $n=1$ and $U=4$, an exact energy can be obtained 
with PBC: $e(1)=-0.8618(2)$ \cite{Becca2000,Sorella}, which 
is below our result: $-0.8559(4)$ \cite{note_fitvsrealn=1}.
Since our largest systematic error is expected to occur here (maximum $n$),
this suggests that the tendency for PS would, if anything,
be {\em underestimated\/} by our energies.
Under TABC the entire density range (including half-filling) is 
treated with the same  approach. 
All calculations use the corresponding
free-electron wave function as
$|\Psi_T\rangle$.
An identical
procedure is applied which has 
no tuning or adjustable parameters.
Clearly the constraining $|\Psi_T\rangle$ has no minimum in $e_h$, 
but an unambiguous minimum emerges from the calculations.
Neither does $|\Psi_T\rangle$ contain spin order, but
the AF ordering appears and vanishes, consistently with the
behavior of the energy.

In summary, recent advances in QMC techniques have enabled us to 
determine the equation of state numerically 
in the 2-D 
Hubbard model at intermediate interactions. 
Our results show that, upon doping, the ground state separates
into one phase
with AF order (hole-free) and the rest without 
($n_c\sim 0.92$ for $U=4$). 
(The nature of the spatially inhomogeneous state 
will require further investigation, for example, the distinction 
between a phase-separated state in finite lattices and 
density waves with long wavelengths, as discussed 
in Sec.~III.A.
More calculations are on-going, which
we plan to report in a future publication.)
The size of the AF spin-density wave region vanishes at $n_c$, causing  the
system to lose long-range AF order.

\section{Acknowledgement}

This work was supported by ARO (No.~48752PH). 
SZ also acknowledges support from NSF (DMR-0535592).
We thank S.~Sorella for sending us the $1/2-$filling data, and
E.J.~Walter for help with computing.
Computations were carried out 
at CPD and the SciClone Cluster (W\&M), and at NCSA.

\end{document}